\documentclass[prb,a4paper,twocolumn,floatfix,showpacs,showkeys,amsmath,amssymb,nobibnotes,altaffilletter]{revtex4}

\usepackage{graphicx}
\usepackage{pslatex}
\usepackage{xspace}

\usepackage{subfigure}
\usepackage{xspace}

\newcommand{\CIGSSe}{Cu(In,Ga)(S,Se)$_2$\xspace}

\newcommand{\CISe}{CuInSe$_2$\xspace}
\newcommand{\degree}{$^{\circ}$}

\begin{document}

\title{Spectroscopy of Electronic Defect States in \CIGSSe-based
Heterojunctions and Schottky Diodes under Damp Heat Exposure}

\author{C.~Deibel}
\author{V.~Dyakonov}
\author{J.~Parisi}\email{Parisi@ehf.uni-oldenburg.de}
\affiliation{Department of Energy and Semiconductor Research,
Faculty of Physics, University of Oldenburg, 26111 Oldenburg, Germany}

\begin{abstract}
The changes of defect characteristics induced by accelerated lifetime tests on
the heterostructure n-ZnO/i-ZnO/CdS/\CIGSSe/Mo relevant for photovoltaic energy
conversion are investigated. We subject heterojunction and Schottky devices to extended damp heat exposure at 85{\degree}C ambient temperature and 85\% relative humidity for various time periods. In order to understand the origin of the pronounced changes of the devices, we apply current--voltage and capacitance--voltage measurements,
admittance spectroscopy, and deep-level transient spectroscopy. The fill factor
and open-circuit voltage of test devices are reduced after prolonged damp heat
treatment, leading to a reduced energy conversion efficiency. We observe the
presence of defect states in the vicinity of the CdS/chalcopyrite interface. Their activation energy increases due to damp heat exposure, indicating a reduced band bending
at the \CIGSSe surface. The Fermi-level pinning at the buffer/chalcopyrite interface, maintaining a high band bending in as-grown cells, is lifted due to the damp-heat exposure.  We also observe changes in the bulk defect spectra due to the damp-heat treatment.
\end{abstract}



\maketitle


\CIGSSe-based thin film solar cells display relatively high energy conversion
efficiencies on laboratory scale devices~\cite{contreras99} as well as large
area modules~\cite{palm02,powalla02}. These heterostructures basically consist of a transparent front contact (highly doped ZnO), a buffer layer (usually CdS), the \CIGSSe chalcopyrite absorber itself, and a back contact (e.g., Mo). An important key prerequisite for the success of this type of solar cells as a powerful
source of renewable energy is their long-term stability under outdoor
conditions. Encapsulated \CIGSSe based thin-film modules have been demonstrated to
perpetuate their performance over many years under various environmental
conditions, including the well-established damp heat (DH) test~\cite{karg97}. Non-encapsulated \CIGSSe-based solar cells, however, show losses in the fill factor and open-circuit voltage after DH exposure, whereas the short-circuit current remains almost unaffected. 
In the present study, we analyze the influence of accelerated lifetime tests on
non-encapsulated ZnO/CdS/\CIGSSe/Mo heterostructure solar cells and
Cr/\CIGSSe/Mo Schottky contacts, focussing on the electronic properties of the
\CIGSSe absorber and the interface between window and absorber layer.


The samples investigated consist of non-encapsulated ZnO/CdS/\CIGSSe/Mo
heterojunction solar cells and Cr/\CIGSSe Schottky
contacts. The chalcopyrite absorber films (thickness about 1.5$\mu$m) was fabricated via a two-step process based on rapid thermal processing of stacked elemental layers. A 20nm thick buffer layer of CdS was deposited on top of the \CIGSSe film in a chemical bath. The transparent front contact, a ZnO layer of about 800nm thickness, is deposited by r.f. and d.c. sputtering. The Schottky devices were processed by deposition of 50nm thick Cr film on top of the \CIGSSe film, followed by a 200nm thick Au layer. Accelerated
lifetime testing was performed under the standardized DH conditions at
85{\degree}C ambient temperature and 85\% relative humidity for various time
periods (24h, 144h, 294h, and 438h). The Schottky devices were exposed to the DH test only for 24h (before Cr/Au evaporation), as the effect of heat and humidity takes place on shorter time scales due to the lack of a protecting ZnO window layer.
We have characterized the modification of the solar cell parameters of our
samples using current--voltage measurements under AM 1.5 (solar spectrum)
illumination at 25{\degree}C. Capacitance--voltage and admittance spectroscopy were
performed using a Solartron 1260 impedance analyzer, deep-level transient spectroscopy (DLTS) by either applying a Semitrap 82E spectrometer or a custom-built transient DLTS setup based on a Boonton 7200 capacitance meter (response time about 120$\mu$s).  The Solartron impedance analyzer was operated with an alternating voltage of 30mV amplitude at frequencies in the range between 1Hz and 1MHz. The Semitrap spectrometer applies a fixed alternating voltage amplitude of 100mV, for the transient-DLTS setup we chose 50mV. In both cases, the sampling
frequency for measuring the transients of the sample capacitance is 1MHz. The DLTS experiments typically included measurements with and without minority-carrier injection. In order to achieve the latter, filling pulses of 1.5V height and $100\mu$s length superimposed on a quiescent reverse bias of -1.5V were applied. For minority-carrier injection, 0.5V forward bias pulses (superimposed on zero bias) were used. The transients were evaluated with exponential fits or the Laplace transform method~\cite{dobaczewski94}. Temperature-dependent analysis was performed using a helium closed-cycle cryostat. For this study, we have characterized about 50 samples. Most of the devices originated from the same process run, otherwise reference samples have been used to ensure comparability.



\begin{figure}
		\centering
			\subfigure{\includegraphics[height=6.5cm]{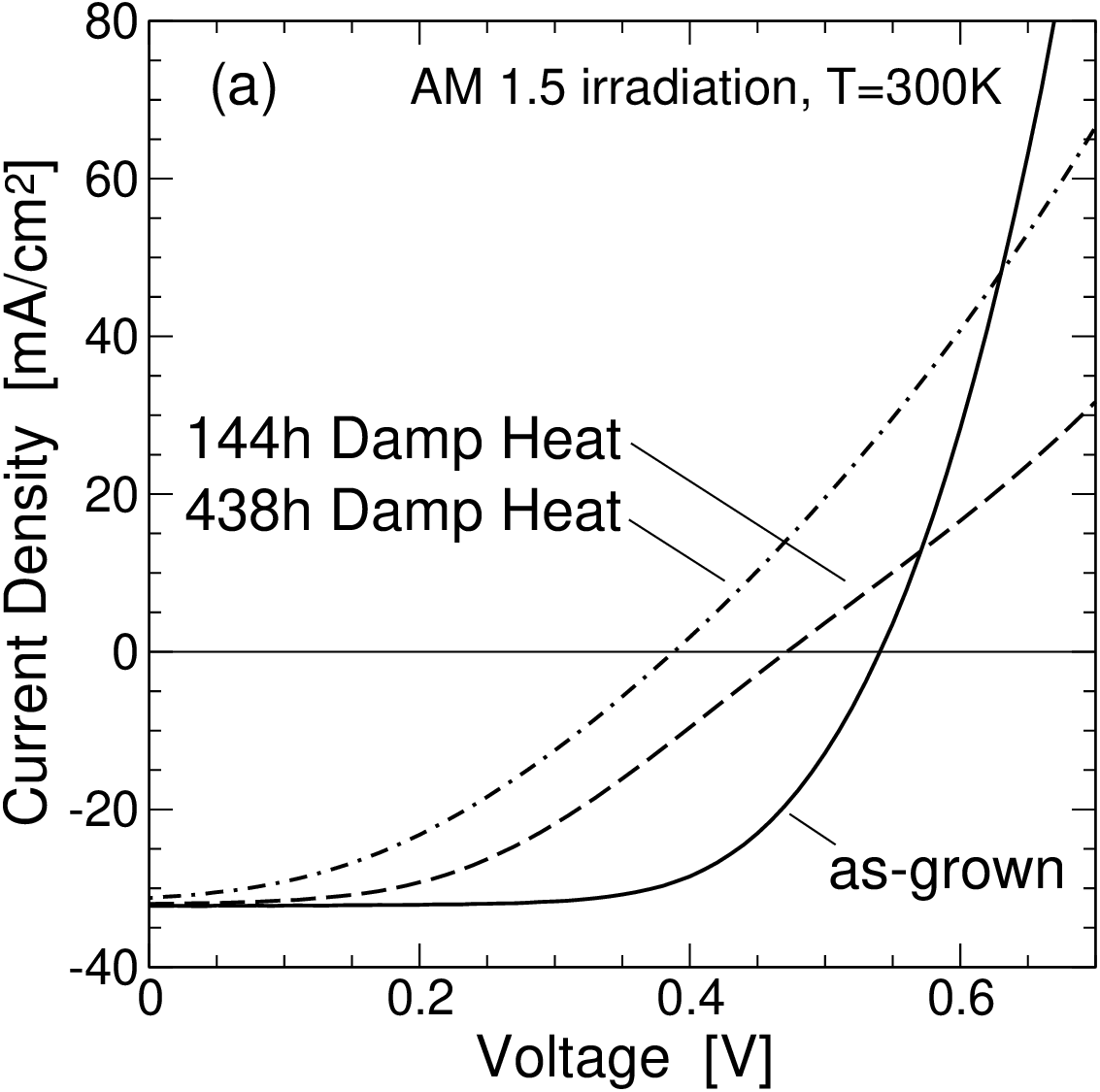}}
			\subfigure{\includegraphics[height=6.5cm]{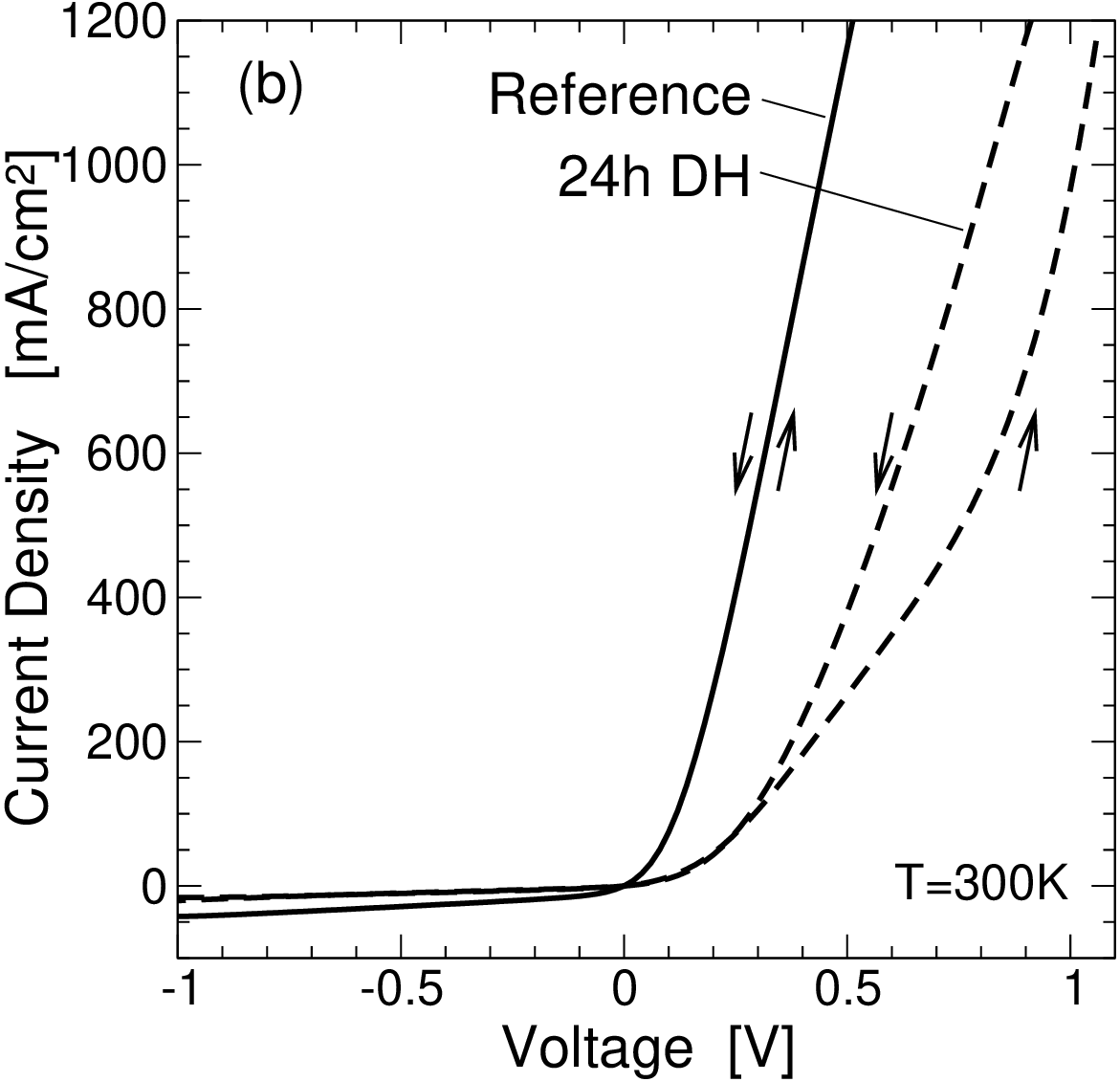}}
		\caption{Current--voltage characteristics of (a) \CIGSSe-based
				heterojunction solar cells under AM 1.5 illumination and (b)
				Cr/\CIGSSe Schottky contacts. The hysteresis is also observed
				in DH treated solar cells.}
		\label{fig:iv}
	\end{figure}

The effect of exposing our non-encapsulated test samples to water vapor manifests in
losses in the fill factor and the open-circuit voltage (see fig.~\ref{fig:iv}(a)). A slight kink in the current--voltage characteristics of the stressed devices can indicate an additional barrier being developed due to the DH exposure. However, we were not able to clearly show the existance of a second space charge region using admittance spectroscopy. A hysteresis of the current--voltage characteristics at forward bias can only be observed for the DH treated devices when the direction of the voltage sweep is changed. This effect is also seen in the DH treated Schottky contacts, as shown in fig.~\ref{fig:iv}(b).

The effective doping density of the \CIGSSe based hetero\-structure samples was
determined using capacitance--voltage ($C$--$V$) measurements. These
experiments were performed at 90K temperature and 100kHz modulation frequency, in order to minimize the influence of deep levels. The slope of the resulting Mott-Schottky plots ($1/C^2$ versus $V$) clearly shows that the net doping concentration diminishes from $6\times10^{15}$cm$^{-3}$ for as-grown cells to about $2\times10^{15}$cm$^{-3}$ for cells after 144h and 438h of DH treatment. However, the built-in potential determined by this method is generally overestimated by approximately a factor of three.


A defect state at the CdS/\CIGSSe interface, further on called $\beta$, can be observed using admittance spectroscopy or DLTS in heterostructure cells~\cite{herberholz98} and Schottky contacts~\cite{deibel03}. The properties of $\beta$ determined by capacitance spectroscopy show that is is a defect state at the surface of the absorber layer (facing the buffer layer)~\cite{herberholz98,deibel03}. Thus, its activation energy represents the position of the Fermi level at the buffer/absorber interface relative to the conduction band~\cite{herberholz98}. A more detailed discussion of the facts indicating that $\beta$ is located at the buffer/absorber interface is given in~\cite{deibel03}. Cells exposed to the DH environment show a continuous shift of the activation energy and capture cross-section of the interface state $\beta$ proportional to the exposure time. The main part of this shift proves to be irreversible.  $\beta$ is probably related to (S,Se) vacancies (dangling bonds) at this interface, which could also explain its sensitivity to oxygen~\cite{kronik00}. As-grown \CIGSSe-based solar cells usually show Fermi-level pinning at the buffer/absorber interface~\cite{herberholz98,deibel01}. The influence of DH treatment on devices that are exposed to the test conditions with the complete heterostructure gives rise to an unpinning of the Fermi level, which manifests in a shift of the interface state $\beta$ when applying an external bias voltage. Devices containing DH treated absorbers, however, still show Fermi-level pinning.

\begin{figure}
	\subfigure{\includegraphics[height=6.5cm]{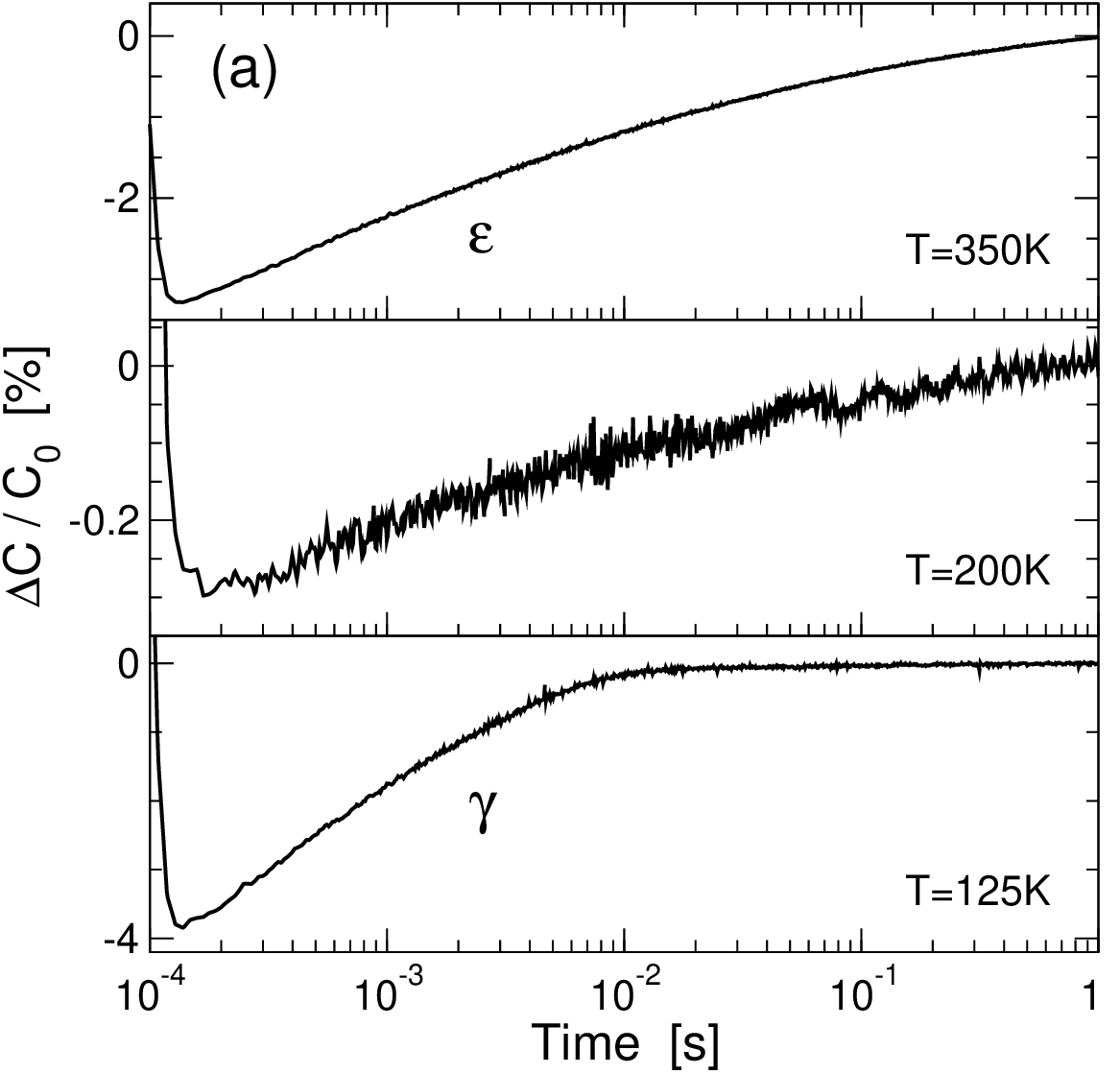}}
		\subfigure{\includegraphics[height=6.5cm]{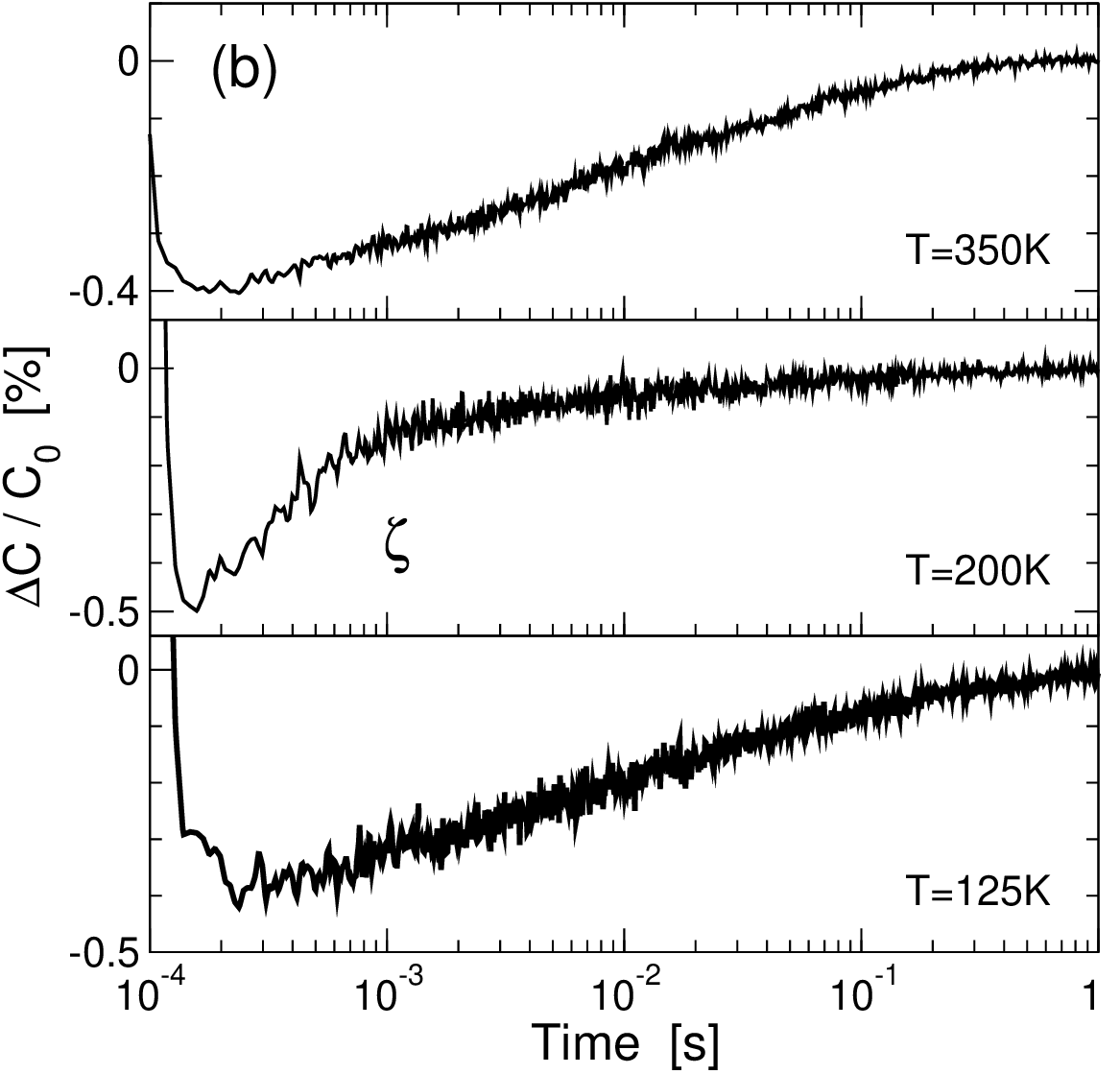}}
	\caption{Capacitance transients of (a) an as-grown and (b) DH-treated (144h) \CIGSSe 				heterojunction
			sample at different temperatures. The 1.5V filling pulse of
			$100\mu$s length was superimposed on a quiescent reverse bias of
			\mbox{-1.5V}.
	\label{fig:transiente-1e-4}}
\end{figure}

A deep defect state, referred to as $\varepsilon$, is observed in DLTS spectra recorded before and after DH exposure. It is located in the absorber bulk. We observe minority and majority DLTS signals (depending on the amount of injected minority carriers
during the filling pulse), both of them resulting in activation energies of
about 550meV. Consequently, we expect the defect state $\varepsilon$ to be a
recombination center. Its capacitance transient at 350K for the case without
injection pulses can be observed in fig.~\ref{fig:transiente-1e-4}(a) for
as-grown samples and in fig.~\ref{fig:transiente-1e-4}(b) for cells exposed
to DH conditions for 144h. The normalized amplitude $\Delta C/C_0$ of the
capacitance transient related to $\varepsilon$ is diminished due to the DH
treatment. The origin of the deep defect state $\varepsilon$ is currently unknown.

We observe a bulk acceptor-like defect state, $\gamma$, with DLTS in as-grown samples only. The corresponding capacitance
transient at $T=125$K is shown in fig.~\ref{fig:transiente-1e-4}(a). We
determine an activation energy of about 160meV. Similar to $\varepsilon$, the
normalized amplitude $\Delta C/C_0$ is diminished due to the DH treatment. Relating the determined defect activation energy to the calculations by Zhang et al.\cite{zhang98}, the defect state $\gamma$ could possibly be an In vacancy.

A deep acceptor state, called $\zeta$, is detected with admittance spectroscopy and
DLTS in as-grown and DH treated Schottky contacts as well as in samples
containing absorbers exposed to DH conditions. In samples DH treated as
complete heterostructure, we observe $\zeta$ with DLTS with a relatively small amplitude, nevertheless manifesting an increase of its concentration relative to as-grown samples.
The defect state corresponds to the fast decay of the capacitance transient of
the DH treated sample at $T=200$K, as shown in
fig.~\ref{fig:transiente-1e-4}(b). From the
admittance data, we obtain an activation energy of about 380meV for $\zeta$.
Note that in co-evaporated \CISe-based cells, the defect state $\zeta$, also
referred to as N2, is present with high concentration already in the as-grown
state, such that it can easily be detected using admittance
spectroscopy. In that cell type, its concentration is
proportional to the time elapsed under DH conditions~\cite{schmidt00}. The defect level $\zeta$ is probably intrinsic~\cite{turcu02}. Taking theoretical considerations into account~\cite{zhang98}, $\zeta$ might be a Cu$_\textrm{In}$ antisite. 

\begin{figure}
		\includegraphics[height=7cm]{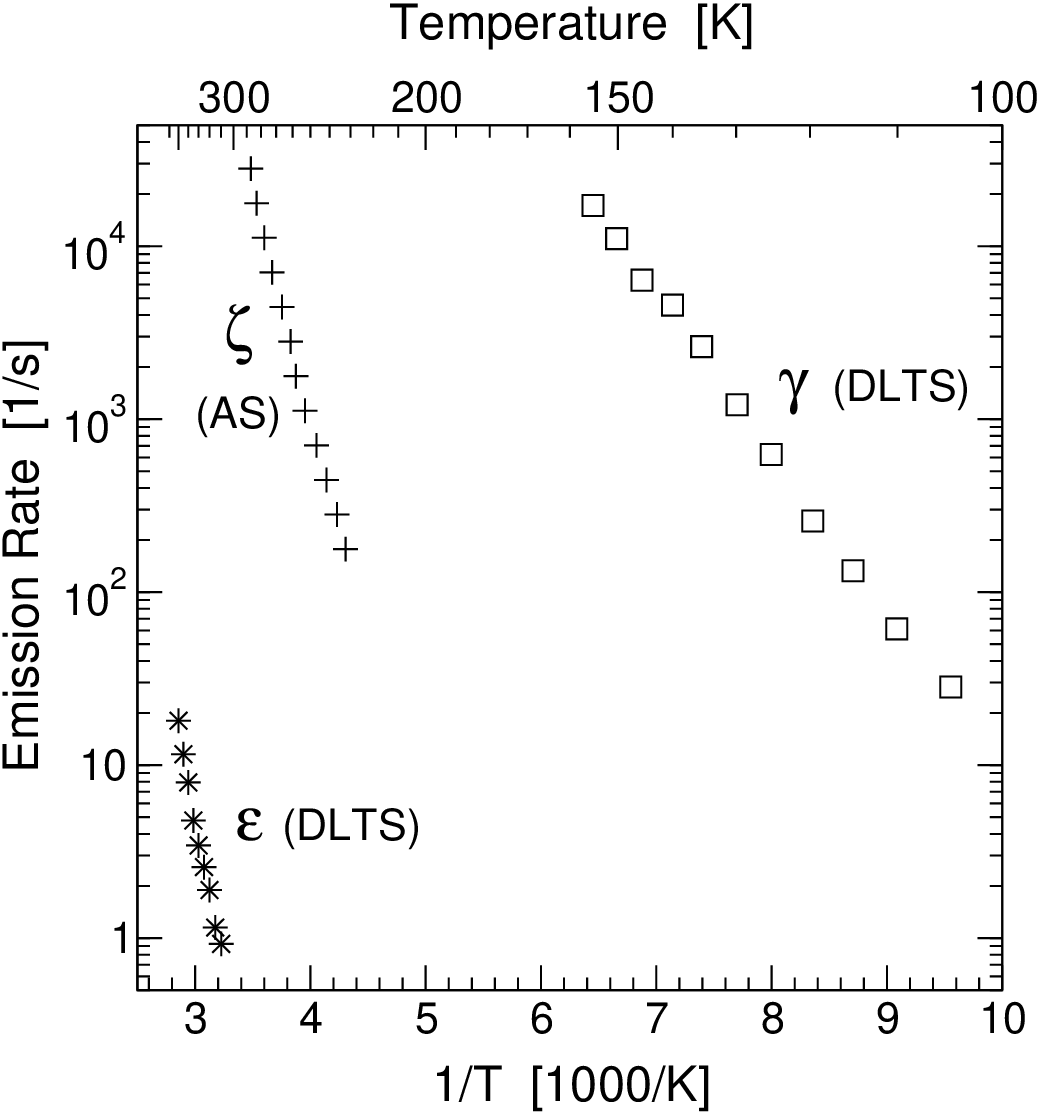}%
		\caption{Arrhenius plot of the three defect states $\gamma$ (acceptor),
			$\zeta$ (acceptor),  and
			$\varepsilon$ (recombination center). The abbreviation AS stands for
			admittance spectroscopy.
		\label{fig:arrh-dh-bulk}}
	\end{figure}

The temperature-dependent emission rates of the three bulk defect
states are displayed in the
Arrhenius plot, fig.~\ref{fig:arrh-dh-bulk}. For the deep trap
$\varepsilon$, only the majority-carrier response is presented. We are not able to present the quantitative changes of the trap concentrations, as their determination is complicated by
several circumstances: First, the transient amplitude strongly depends on
temperature, indicating the influence of a temperature-dependent capture
cross-section and leakage currents (especially for the DH treated device) on the
amplitude~\cite{chen84}. Also, the transient is non-exponential, and both amplitude and decay strongly depend on the width of the filling pulse. The qualitative trap concentrations of the two defect states $\gamma$ and $\epsilon$ decrease due to the DH treatment, whereas the trap level $\zeta$ shows a concentration increase. 

Our experimental findings can only be seen as a starting point for an identification of the causes of DH induced solar cell degradation. On one hand, further study is needed to gain more insight into the correlation of electronic and chemical modifications. On the other hand, a quantification of the defect state concentrations is crucial for modelling the implications of the DH treatment on the electronic device properties.


To summarize our experimental results, we observed major changes of the \CIGSSe absorber layer and the CdS/\CIGSSe interface due to damp-heat treatment of \CIGSSe-based solar cells and Schottky contacts. We found a reduced band bending and a decreased net doping density in the chalcopyrite absorber layer. Three absorber bulk traps with non-exponential
capacitance transients were detected, qualitatively their net concentration decreases due to the damp-heat treatment. The observation of a hysteresis in the current--voltage
characteristics of damp-heat treated heterojunctions and Schottky contacts
indicates an increased concentration of traps in the absorber layer. The
Fermi-level pinning at the buffer/chalcopyrite interface, maintaining a high
band bending in as-grown cells, is lifted due to the damp-heat exposure. The combination of these effects represent a major part of the damp-heat induced device degradation.

\acknowledgments
The authors would like to thank J.~Palm and F.~Karg (Shell Solar, Munich) for
interesting discussions and for providing the heterojunction samples and
absorber layers. Fruitful discussions with M.~Igalson and the partners of the
Shell Solar joint research project at the University of W{\"u}rzburg and the
Hahn Meitner Institute Berlin are also acknowledged. 


\end{document}